\documentclass[runningheads,citeauthoryear]{apinv}
\usepackage{epsfig,cite,graphics}
\usepackage[T2A]{fontenc}
\usepackage[cp1251]{inputenc}
\usepackage[bulgarian,english]{babel}

\begin{document}
\setcounter{page}{1}
\title{Intra-night flickering of RS Ophiuchi: \\
I. Sizes and cumulative energies of time structures }
\author{Ts. B. Georgiev\inst{1},
R. K. Zamanov\inst{1}, 
S. Boeva\inst{1}, 
G. Latev\inst{1},
B. Spassov\inst{1}, \\  
J. Mart\'{\i}\inst{2}, 
G. Nikolov\inst{1}, 
S. Ibryamov\inst{3},  
S. V. Tsvetkova\inst{1},  
K. A. Stoyanov\inst{1}}
\titlerunning{Intra-night flickering of RS Oph}
\authorrunning{Ts. B. Georgiev et al.}
\tocauthor{Ts. B. Georgiev,
R. K. Zamanov, 
S. Boeva, 
G. Y. Latev,
B. Spassov, 
J. Mart\'{\i}, 
G. Nikolov, 
S. Ibryamov,  
S. V. Tsvetkova,  
K. A. Stoyanov }

\institute {Institute of Astronomy and National Astronomical Observatory, \\ Bulgarian Academy of Sciences, 72 Tsarigradsko Chaussee Blvd., 
	  1784 Sofia, Bulgaria
	\and Departamento de F\'{\i}sika (EPSJ), Universidad de Ja\'{e}n,\\ Campus Las Lagunillas  A3-420, 23071 Ja\'{e}n, Spain
	\and University of Shumen, 115, Universitetska Str., 9700 Shumen, Bulgaria
\newline \email{tsgeorg@astro.bas.bg }}
\papertype{Submitted on XX.XX.2018; Accepted on XX.XX.2018}	
\maketitle

\begin{abstract}
We analyzed 29 pairs of time series in B and V bands of the recurrent nova RS Oph. 
The observations were carried out in 2008-2017 with duration 0.6–3.6 hours, with time resolution 0.5–3.3 min.  
We scanned digitally each series by data windows with various sizes $\Theta$ and derived two of the simplest fractal 
parameters for every $\Theta$  –  standard deviation $D$ and structural deviation $S$.
Using the local minima of the structural function log $S$ = $f_S$(log $\Theta$) we unveiled 80 time structures, 
42 in B band and 38 in V band, with time sizes 10-120 min. 
About 3/4 of the time sizes belong to the interval 10-40 min and about 1/4 lie in the interval 60-120 min. 
The respective cycles per day are 144-36 c/d and 24–15 c/d. On logarithmic scale, the distribution of the time 
sizes shows maximums at about 10, 21, 36 and 74 min. The 10 min flickering is poorly detectable in our series and 
we found the most widespread time structures (in about 1/5 of the  cases) have time sizes about 21 min (about 69 c/d).
Using the deviation function log $D$ = $f_D$(log $\Theta$) we estimated the relative cumulative energy (including the energy of the shorter structures in it), associated with the detected structure sizes,   
to be in the interval of the relative fluxes 2-11 \%. The energies correlate weakly with the logarithms of the structure sizes,  with correlation coefficients 0.60 and 0.57,  
under slope coefficients 0.04 and 0.03 in B and V band, respectively. The distributions of the energies occur bimodal, 
with maximums about 4 \%  and 6 \% in B band, as well as about 3\% and 5\% in V band. 
The left and right modes of the distributions may be associated with the structure sizes 10 – 21 min and 37 – 74 min, respectively. 

\end{abstract}
\keywords{stars: binaries: symbiotic -- novae, cataclysmic variables -- accretion, accretion discs -- stars: individual: RS Oph }

\section*{Introduction}

RS~Oph is a symbiotic recurrent nova that contains an M2~III mass donor (Shenavrin, Taranova \& Nadzhip 2011) and a massive carbon-oxygen white dwarf (Mikolajewska
\& Shara 2017). The orbit of the system is circular (Fekel et al. 2000) with period 453.6$\pm$0.4~d (Brandi et al. 2009). RS~Oph exhibits 
recurrent nova outbursts approximately every 20 years (Evans et al. 2008) with most recent nova outburst that occurred on 2006 February 12 
(Narumi et al. 2006). Wynn (2008) proposed that both Roche lobe overflow and stellar wind capture are possible accretion scenarios in the case of RS~Oph.

The flickering (short-term brightness variability in time scales from minutes to hours) of RS~Oph has been detected by Walker (1977). 
The peak-to-peak  amplitudes of these variations reach 0.3 -- 0.5 mag. The flickering parameters vary from night to night. 
The study of the physics of the flickering requires revealing of the details of the variability and characterizing them by their typical time sizes, amplitudes, standard deviations, morphologies, etc. Such information from the apparent chaos of the flickering may be extracted in different ways, as well as by conventional statistical methods, considered today as elements of the fractal analysis. 

The fractal approach is preferable here because (i) it is conceptually simple, and (ii) it is weakly sensitive to non-equality of the data sampling. Similar approach has been applied by Bachev et al. (2011) and Georgiev et al. (2012) for studying of the flickering of the cataclysmic variable star KR~Aur. The main result there is that the flickering contains at least two different sources of variability. 

Recently, Kundra,  Hric \& G\'{a}lis (2010) carried out wavelet analysis of two time series of RS~Oph. They unveiled two different modes of flickering, with frequency 50-100 c/d and $< 50$ c/d (cycles of variations per day). The respective quasi-periods are 30-15 min or $> 30$ min. The amplitudes of the flickering modes are estimated to be about 0.1 mag or about 0.6 mag, respectively. Later, Kundra \& Hric (2014) revealed two flickering modes with frequency 60 c/d and 140 c/d, i.e. with quasi-periods 24 min and 10 min. The amplitudes are 0.6 mag and 0.1 mag, respectively. The flickering phenomenon is explained (yet roughly) by variable mass transfer from the red giant through the accretion disk to the surface of the white dwarf. The reasons of the appearance of two modes of flickering are not clear. 

In the present paper we concentrate on the revealing of time structures with sizes 10 -- 120 min and on the estimation of their cumulative energies. Section 1 presents the observing material. Section 2 describes the method of data processing. Section 3 shows some examples. Section 4 presents the results. Section 5 contains the conclusions.

\section*{1. Observations} 

We analyzed 29 pairs of light curves of RS Oph in B and V bands. These are the observations 
analyzed in Zamanov et al. (2018) to which  20090614 is added. 
The observations were carried out with the 2.0m RCC, the 60cm Cassegrain and
the 50/70cm Schmidt telescopes of the Rozhen NAO, the 60cm Cassegrain telescope of the Belogradchik Astronomical Observatory, 
as well as the 41cm telescope of the University of Ja\'{e}n, Spain, in 2008-2017. The magnitude standard error is 0.005 – 0.01 mag.
The data about the observing material are collected in Table 1 and Table 2. 

The duration of a single monitoring is $T_M$ = 0.6 – 3.6 h, the number of the data points (CCD frames) is $N_P$ = 40 – 470 and the 
mean time resolution is $\delta T$ = 0.3 – 3.3 min. The total average magnitudes of RS Oph are: B  = 12.17 $\pm$ 0.37 mag (from 11.43 to 12.90 mag), 
V = 11.06 $\pm$ 0.38 mag (from 10.38 to  11.65) and (B-V) = 1.10 $\pm$ 0.09 mag (from 0.097 to 1.26 mag). 

Our processing is based on standard and structural functions (Eq.4). It requires approximative flat time series, i.e.time series in which the local structures appear among almost constant general mean value. However, most of the flickering light curves look like parts of curves of magnitude variations with time scales of a few hours (Fig.3(b)) or 
they are significantly biased (Fig.6). About 1/3 of all series appeared not susceptible to direct processing. By this reason we were forced to apply at least linear unbiasing (flattening) of all series. 

We fitted the original magnitude series $m’(t)$ by linear regressions $m’(t) = G_0 + G.t$ and later we processed only the unbiased magnitudes $m(t)$: \\
\begin{equation}
 m(t) = m’(t) - G_0 - G.t + m’_0, 
\end{equation}
where $m’_0$ is the average magnitude of the original series which is added to keep the mean value of the original data.

After this flattening all series became good for our processing. In respect to the original series $m’(t)$ the magnitude standard deviations of the unbiased series $m(t)$ decrease about 1.3 times (from 1.0 to 2.5  times). They became 0.062 $\pm$ 0.018 mag in B band and 0.050 $\pm$ 0.016 mag in V band. The respective amplitudes (full ranges of variations) become 0.15 – 0.45 mag, with average 0.28 $\pm$ 0.07 mag in B band, as well as  0.12 – 0.40 mag,
 with average 0.25 $\pm$ 0.07 mag, in V band (Table 1).

As a result of the flattening procedure some information about the long time magnitude variations is lost, 
but the structures in the flickering with time sizes 40 – 80 min can be better revealed. However, the bias gradients (slope coefficients) $G$ (Eq. 1) contain some information about the magnitude variations on the scale of a few hours. (Preliminary we made ourselves sure that $G$ does not depend on the time in the period 2008-2017.) 

Figure 1(\textit{a}) shows that the bias gradient $G$ anti-correlates very weakly with the average value of the series. The coefficients of correlation are -0.23 and -0.30 in B and V bands, respectively. 
This faint correlation is due to the presence of relatively steeper general increasing of the brightness in some series. The respective points are placed in the bottom part of Fig. 1 and correspond to series with relatively low average values (low sate ofRS Oph). 

\begin{figure}[!htb]  \begin{center}
\centering{\epsfig{file=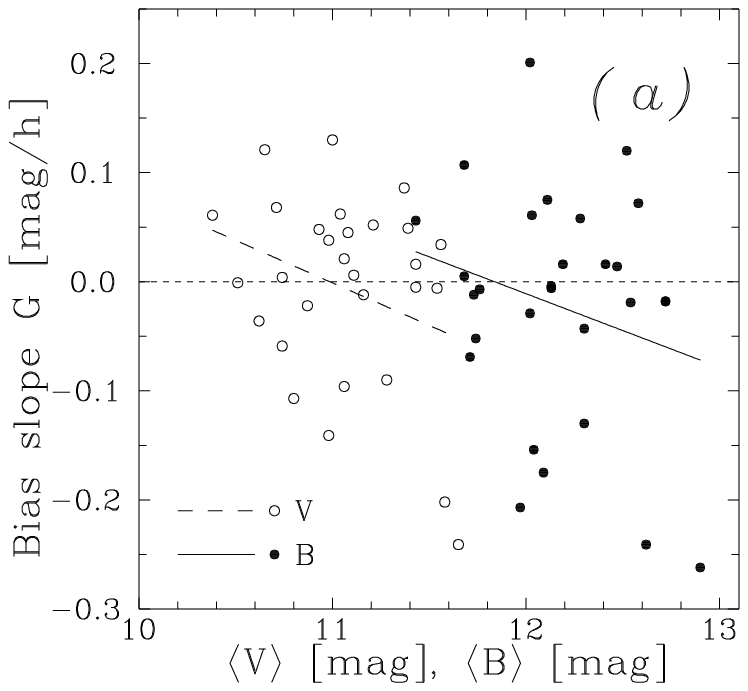, width=.40\textwidth}}
\centering{\epsfig{file=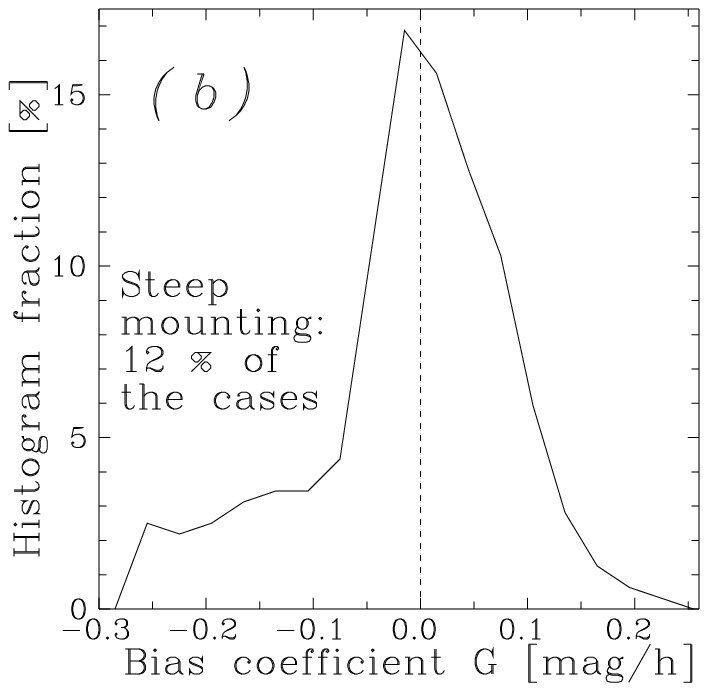, width=.37\textwidth}}
\caption[] {(\textit{a}) Anti-correlations between the average brightness <B> and <V> in the series, in magnitudes,  
and the bias gradient G (Eq. 1), in magnitudes per hour. The regression lines are shown. 
(\textit{b}) Common distribution of the bias gradient $G$ in B and V bands with binning size 0.03. 
(Hereafter the distributions (histograms) are slightly smoothed by convolution with the kernel {0.25, 0.50, 0.25}.} 
\label{fig.1}   \end{center}  \end{figure} 

Figure 1(\textit{b}) shows the common distribution of the gradients $G$ in B and V bands by they values. 
This distribution shows that the mean gradients $G$ (Eq.1) of the monitoring series are 
typically slant, -0.1 < $G$ < 0.1 mag/h. However, about l7/58 (12 \%) of series form well pronounced 
left tail of the distribution with $G$ < -0.15 mag/h. 
The most steep brightness increasing has $G$ = –0.26 mag/h, in the flickering \#12B (Table 2). 
Right tail of the distribution, analogous to the left tail, is not observed. 

\section*{2. Method of analysis}

The fractal conception, systemized conclusively by Mandelbrot (1982), 
is applicable for analysis of 1D, 2D and 3D non regular structures –- from time series to space distribution of the galaxies. The fractal analysis aims to reveal and characterize self-similarity in the apparent chaotic structures. 
Following the recommendations of Russ (1994), Hastings \& Sugihara (1995) and Falkoner (1997) we explore two of the simplest fractal indicators, 
parameters and functions  (Eq.2, Eq.3, Eq.4). 

The fractal analysis of a time series $m(t_n),  n = 1,2, … , N$, is based on a system of data windows with time sizes $\Theta_j ,  j = 1, 2,…, J$. Each window scans the series and takes $k = 1,2, … , K$ different positions among the series. 
	
Figure 2(\textit{a}) illustrates the deriving of the fractal indicators from the $k$-th position of the window with $j$-th size and bounds $t_1$ and $t_2$. The standard deviation $d_{jk}$ 
and the structural deviation  $s_{jk}$ are calculated from the data inside the window as follows:
\begin{equation}
 d_{jk} = m_{sd}; \hspace{8mm} s_{jk} = |m_1 - m_2|/2, \hspace{8mm} k=1,2, … , K. 
\end{equation} 
The first indicator, $d_{jk}$, is the standard deviation of the data in the window. The second indicator, $s_{jk}$, is the structural indicator (see below). 
The division by 2 in  $s_{jk}$ ensures comparability of $s_{jk}$ and $d_{jk}$. 
(In the general case both indicators are used in relative form, i.e. after dividing by the average 
value of the time series inside the window. 
However, by definition the stellar magnitudes are based on relative fluxes and here such division is not necessary.)   

\begin{figure}[!htb]  \begin{center}
\centering{\epsfig{file=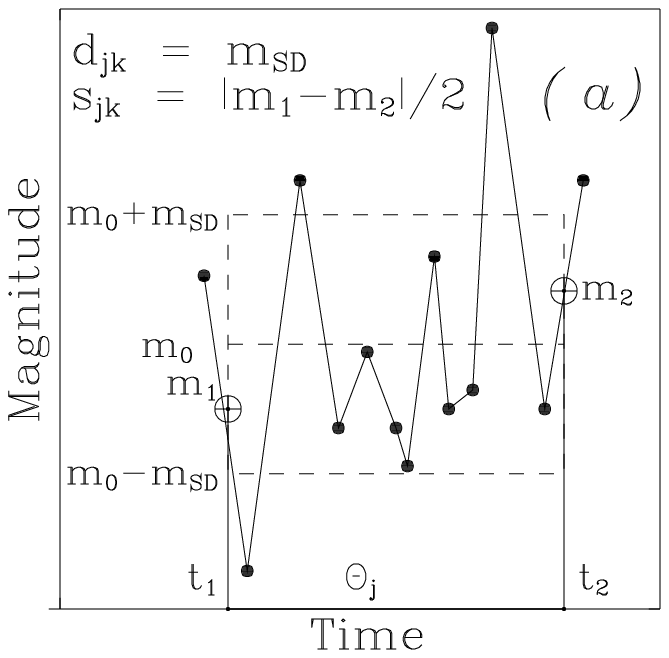, width=.4\textwidth}}
\centering{\epsfig{file=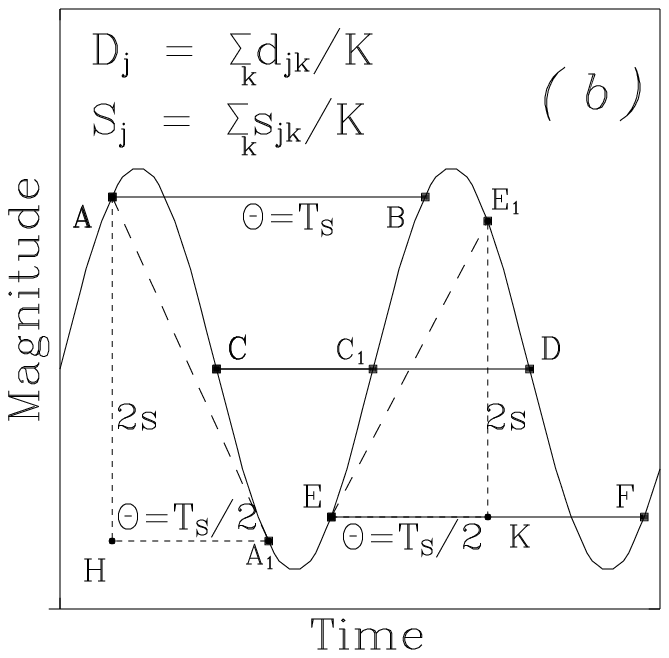, width=.4\textwidth}}
\caption[] {(\textit{a}) Time window with size $\Theta_j$ on its \textit{k}-th position within the time bounds $t_1$ and $t_2$, placed among the flickering data (dots, connected with segments). Formulas for the deriving of the indicators $d_{jk}$  and $s_{jk}$(Eq. 2) are included; (\textit{b}) Account of an ideal sinusoidal shape in the series with period $T_S$ by the structural indicator $s_{jk}$ (see the text). Formulas for deriving of the parameters $D_j$ and $S_j$(Eq. 3) are included. } 
\label{fig.2}   \end{center}  \end{figure} 
               
The average values of the indicators from all $K$ positions of the $j$-th window, with size $\Theta_j$ are defined as fractal parameters -- deviation parameter $D_j$ and structural parameter $S_j$ \\
\begin{equation}
 D_j = \sum d_{jk}/K; \hspace{8mm} S_j = \sum s_{jk}/K; \hspace{8mm}  j=1,2,…, J.
\end{equation} 

The dependence of the fractal parameters $D_j$ and $S_j$ (Eq.3) on the size of the scanning window $\Theta_j$, 
in $\log$-$\log$ coordinates, are the fractal functions:  \\
\begin{equation}
  \log D = f_D(\log \Theta); \hspace {12mm} \log S = f_S(\log \Theta). 
\end{equation} 

The deviation function, $f_D$, describes the change of the deviation parameter $D$ with the increasing of the window size $\Theta$. While $\Theta$ is small and increases it incorporates larger structures with larger variations. 
The $f_D$ grows up. But when the window begins to include dominant structures (periodic or quasi-periodic) the quantity of new larger deviations decreases 
and the $f_D$ tends to plateau (Fig. 3-6, bottom panels, short-dashed curves. See also the deviation function of the solar spot number in Georgiev et al. 2012, Fig.4\textit{a}). 

The level of the plateau of the deviation function $f_D$ characterizes the cumulative energy of the variations, associated with such structure, 
together with its substructures. If the time series is normal on uniform random process, larger structures do not exist and $f_D$ has horizontal behavior (Georgiev et al. 2012, Fig. 2). 
Here the deviation function $f_D$ is used only for uniform estimation of energies $E_S$ of time structures with sizes $T_S$, derived preliminary by the structural function (see below). 

The structural function, $f_S$ (Eq. 4), describes the change of the structural parameter $S$ with the increasing of the window size $\Theta$. While $\Theta$ is small and increases, 
$f_S$ increases too, 
like $f_D$. However, if the window size corresponds to the size of periodic or quasi-periodic structure, the $f_S$ shows minimum. 

Figure 2 (\textit{b}) shows a time series that contains an ideal sinusoidal structure with period (time size) $T_S$. 
If the window size is $\Theta \approx T_S$, numerous window positions, like AB, CD or EF, import almost zero contributions 
$s$  to the structural parameter $S$ (Eq. 3). The $f_S$ reaches local minimum at $\Theta \approx T_S$ (and at $\Theta \approx 2T_S$, 
$\Theta \approx 3T_S$, etc., if the series are enough long). Otherwise, if the window sizes is $\Theta \approx T_S$/2 (or  3$T_S$/2, 5$T_S$/2, etc.), numerous window positions, like  AA$_1$ and EE$_1$, 
import significant contributions $s$ ($s$=AH/2 or $s$=KE$_1$/2) into the structural parameter $S$. In such cases the structural function $f_S$ 
reaches local maximums (Fig. 3-6, down panels, solid curves. See also the structural function of the solar 
spot number in Georgiev et al. 2012, Fig.4\textit{a}).  
Note that if the time series is normal on uniform random process, the structural function $f_D$ has horizontal behaviour (Georgiev et al. 2012, Fig. 2).  

The structural function $f_S$ is sensitive to sizes $T_S$ of time structures with roughly sinusoidal shapes. 
Here the positions of the minimums of  $f_S$ are  used for uniform estimations of sizes of time structures. 
Analogous applications of the $f_S$ are given for example by Di Clemente et al. (1996), Bachev et al. (2011) 
and Gantchev et al. (2017). 
 
So, the window size $\Theta$, which corresponds to a significant minimum of the structural function $f_S$, is estimation of a structure size $T_S$, assumed usually as basic or dominant. If this structure repeats, it should be associated with a quasi-periodic structure, 
even if the basic structure has complicated shape (Fig.6). 
The value of $f_D$, which corresponds to the time size $T_S$, is a measure of the cumulative energy of the flickering $E_S$, associated with the specified structure size $T_S$. For example, on Fig.3(\textit{a,b}) the positions of the minimums of $f_S$, denoted by log$T_S$, point out the values of the $f_D$, denoted by log$E_S$.   

Examples of our processing are shown in Fig.3-6. Results about the time structures and their flickering energies 
are presented in Table~3 and in Fig.8-10.

Table~3 contains the estimations of the structure sizes in minutes, in linear and logarithmic scales, as well as the respective 
cumulative energies in magnitudes. Results in B band and V band are juxtaposed in the lines of the Table 3. 
Every line cover data for one detected time size $T_S$. When the time size is detected in B band, 
but not in V band, the V-part of the string is empty, and vice versa.

\section*{3. Examples}
Figures 3-7 show the examples. The top panels represent the variations of the original magnitudes $m’$, but the analysis is applied
on the  linearly unbiased magnitudes $m$ (Eq. 1). The bottom panels 
show the functions $f_D$ = log $D$ and $f_S$ = log $S$ (Eq.~4), each of them is drown by 60-80 points (window sizes). 
Parameters of interest here are the basic structure size $T_S$ and the cumulative relative energy $E_S$, associated with $T_S$.   

So, the positions of the solid vertical segments in the bottom panels show the structure sizes $T_S$, corresponding to the time structures which are considered as basic or dominant. 
The solid horizontal segments show the levels of the respective energies $E_S$, in magnitudes, containing the contributions of all variations 
with time characteristic sizes $T$ < $T_S$. The dashed vertical segments correspond to repeating structures with sizes $2T_S$, $3T_S$, etc. 

\begin{figure}[!htb]  \begin{center}
\centering{\epsfig{file=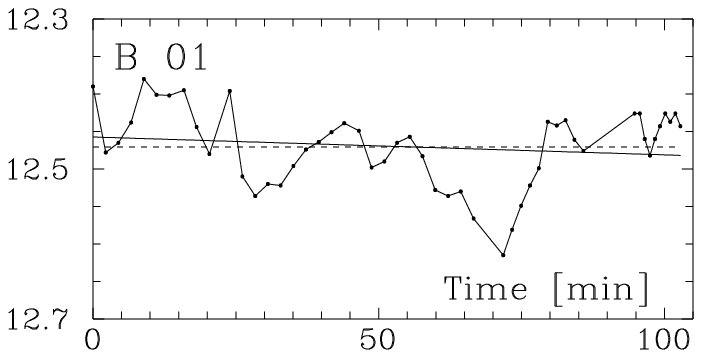, width=.35\textwidth}}
\centering{\epsfig{file=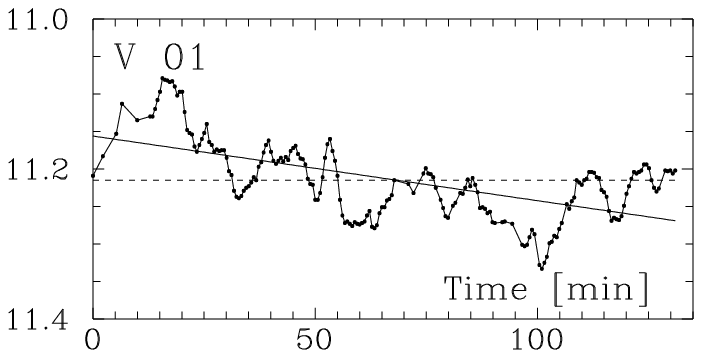, width=.35\textwidth}}
\centering{\epsfig{file=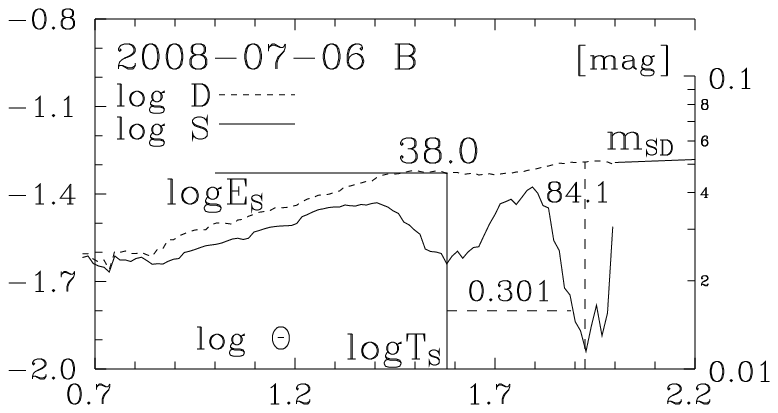, width=.46\textwidth}}
\centering{\epsfig{file=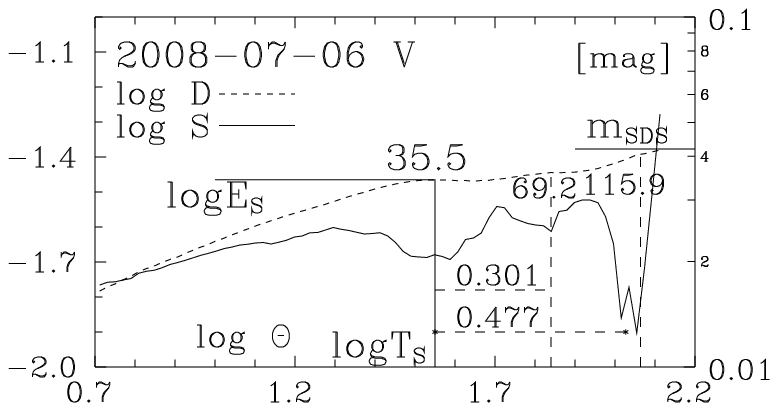, width=.455\textwidth}}
\caption[] { (\#01) 
Top panels: The flickering time series in B and V bands. The dashed lines are the levels of the average magnitudes and the solid lines show the regression lines of the biases. 
Bottom panels: The fractal functions  $f_D$ (dashed curves) and $f_S$ (solid curves). $T_S$ and $E_S$ are the sizes and the energies of the structures, considered as basic. The right ordinates are marked in magnitudes, where the short horizontal segment shows the level of the  standard deviation, $m_{SD}$ of the whole flickering series. } 
\label{fig.3}   
\end{center}  \end{figure} 

Figure 3 represents our first observations in 2008. 
In respect to the beginning of the V band observations, 
the B band observations begin about 40 min later and the large scale biases 
of both series are different. The B band observations has about 3 times lower resolution and it does not show some small details. 
In spite of these circumstances both series are useful. 

Figure 3 reveals a basic structure with sizes $T_S$ = 38 min in B band and $T_S$ = 35.5 min in V band. The respective energies $E_S$, 
expressed in magnitudes, are 0.062 mag and 0.42 mag.
The positions of other $F_S$  minimums, marked by dashed vertical lines, are placed at 0.301 = log 2 or 0.477 = log 3 to the right from the basic minima. 
We neglect them as “harmonics” (2.$T_S$, 3.$T_S$) and self-dependent.

\begin{figure}[!htb]  \begin{center}
\centering{\epsfig{file=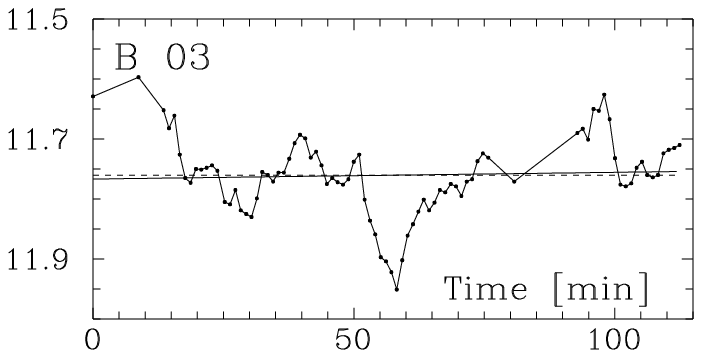, width=.35\textwidth}}
\centering{\epsfig{file=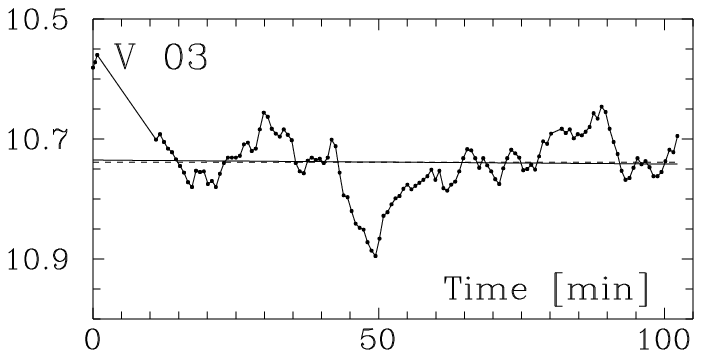, width=.35\textwidth}}
\centering{\epsfig{file=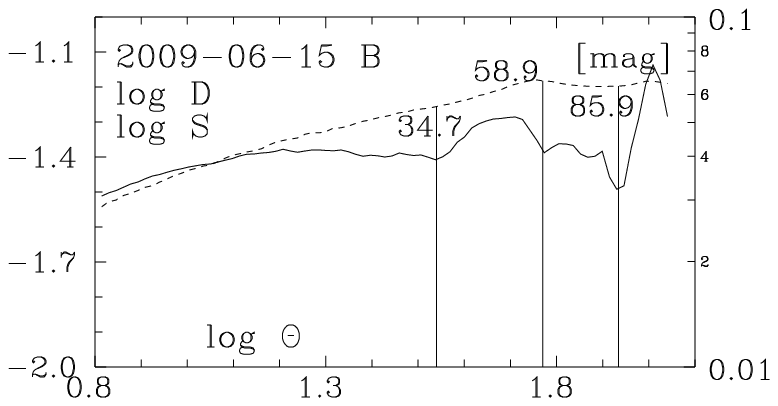, width=.46\textwidth}}
\centering{\epsfig{file=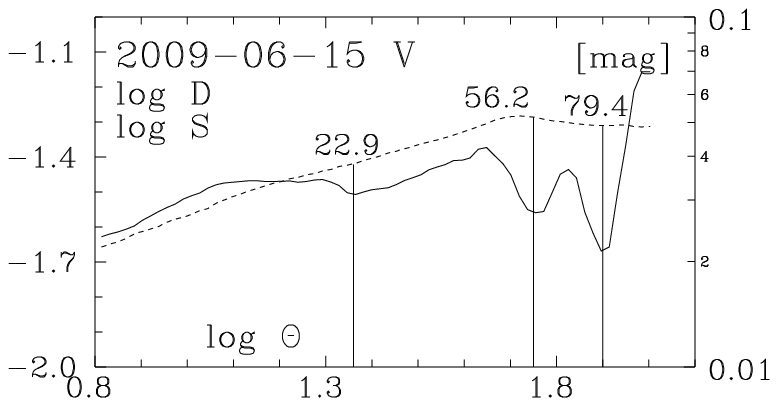, width=.46\textwidth}}
\caption[] {(\#03) The unique case where 3 time structures seem to be self-independent } 
\label{fig.4}   \end{center}  \end{figure} 

Figure 4 represents the unique case with 3 possible independent structure sizes. 

\begin{figure}[!htb]  \begin{center}
\centering{\epsfig{file=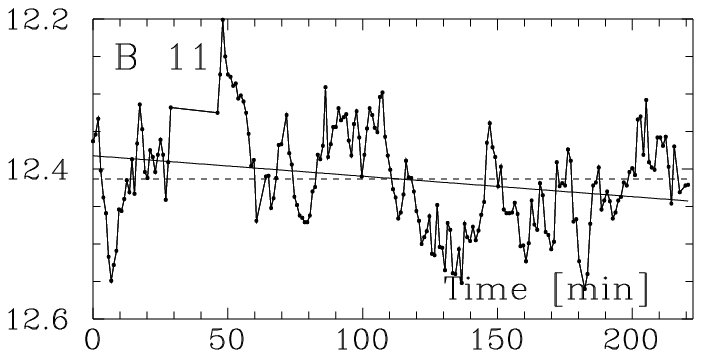, width=.35\textwidth}}
\centering{\epsfig{file=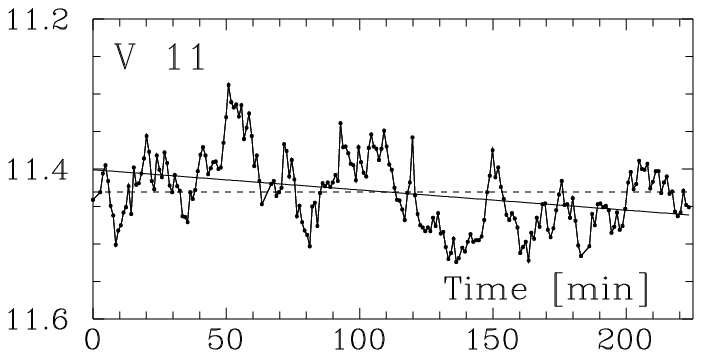, width=.35\textwidth}}
\centering{\epsfig{file=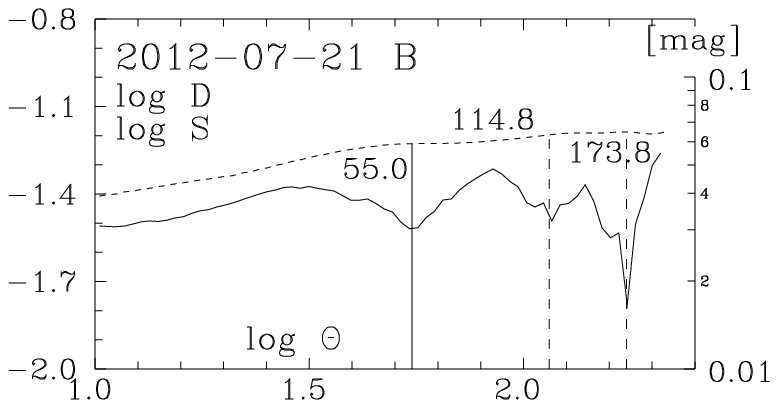, width=.46\textwidth}}
\centering{\epsfig{file=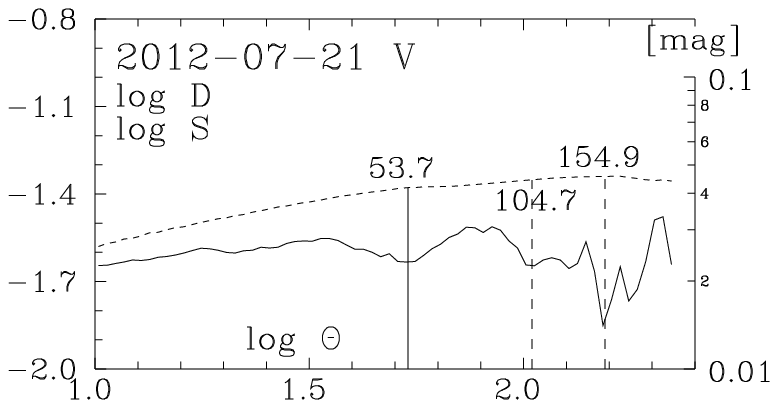, width=.46\textwidth}}
\caption[] {(\#11) The longest light curve where the basic (dominating) structure has size $T_S \approx$ 54 min } 
\label{fig.5}   \end{center}  \end{figure} 

Figure 5 shows the longest observations ($T_M \approx$ 220 min, $N_P$ =220 points). 
It contains the longest detected  structures 
with  sizes $T_S$ = 155 – 174  min. However, these sizes, as well as the sizes with $T_S$ = 105 – 115 min, 
are divisible into the well pronounced shortest size with $T_S$ about 54 min,
multiplied by 2 or 3. The shortest structure size seems to be basic. 

Figure 6 shows a case when the flickering in V band is observed immediately after the flickering in B band, by the same telescope. In both cases the minimums of 
$f_S$ correspond to $T_S$, 2$T_S$, 3$T_S$, 4$T_S$ and even 5$T_S$. These series show a remarkable case where the flickering with $T_S$ about 
21 min is periodic over time interval with common duration about 5 h. The periodicity of the $f_S$ is similar to the periodicity of the $f_S$ of the solar spot number 
(Georgiev et al. 2012, Fig.4). The series in the case \#10 are analogous, again with $T_S \approx$ 21 min.

\begin{figure}[!htb]  \begin{center}
\centering{\epsfig{file=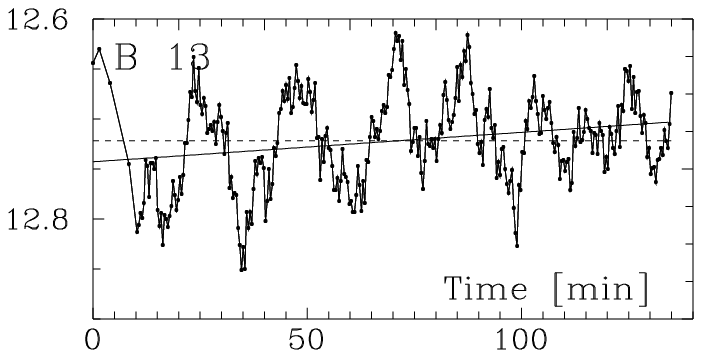, width=.35\textwidth}}
\centering{\epsfig{file=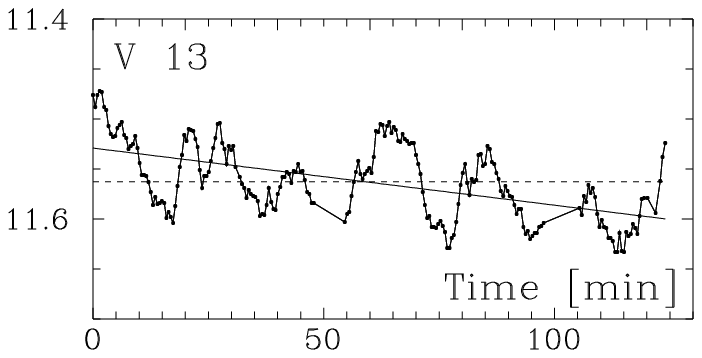, width=.35\textwidth}}
\centering{\epsfig{file=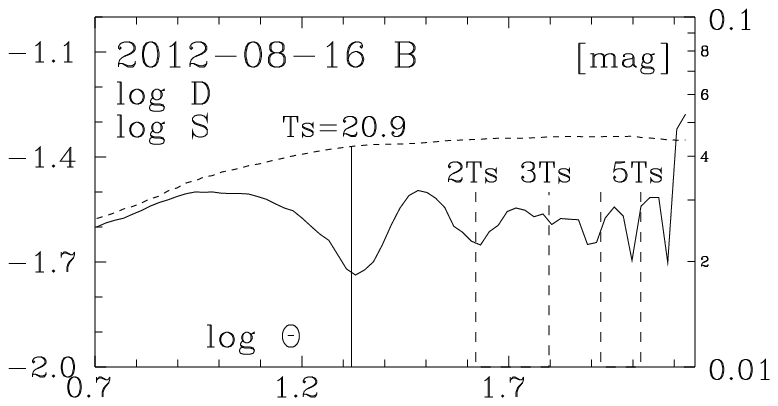, width=.46\textwidth}}
\centering{\epsfig{file=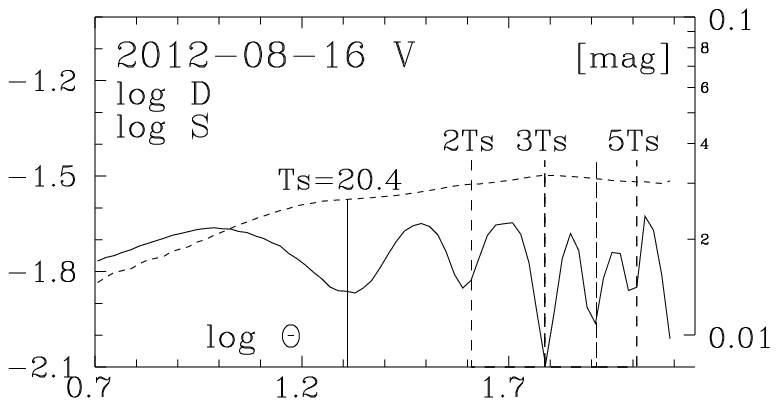, width=.46\textwidth}}
\caption[] {(\#13) One dominant time structure repeats many times in the series. We account only the basic quasi-periods with $T_S$ = 20.9 min in B band and $T_S$ = 20.4 min in V band } 
\label{fig.6}   \end{center}  \end{figure} 

Figure 7 shows as additional example the applications of our method for revealing of characteristic structure sizes 
in the behavior of the average magnitudes and color of RS Oph in the period 2008-2017. 
Because of the small number and bad sampling of the data,  the structural functions 
$f_S$ are almost horizontal and they fluctuate strongly. Though, the time sizes estimations 
in all 3 cases are similar: $T_S \approx$ 8 yr. 
Note that similar time size is hinted (but not estimated) by cubic polynomials (dashed curves). 

\begin{figure}[!htb]  \begin{center}
\centering{\epsfig{file=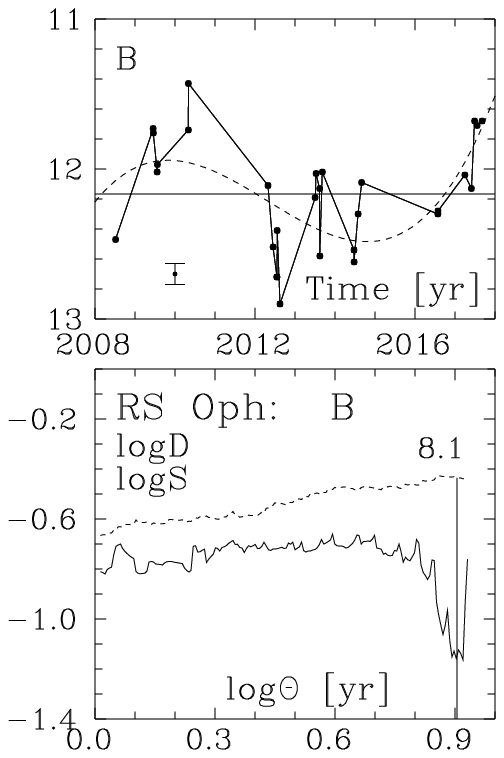, width=.3\textwidth}}
\centering{\epsfig{file=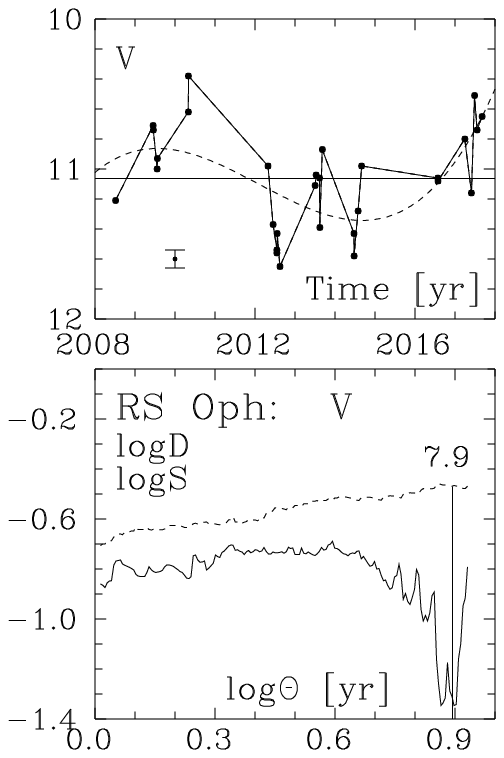, width=.3\textwidth}}
\centering{\epsfig{file=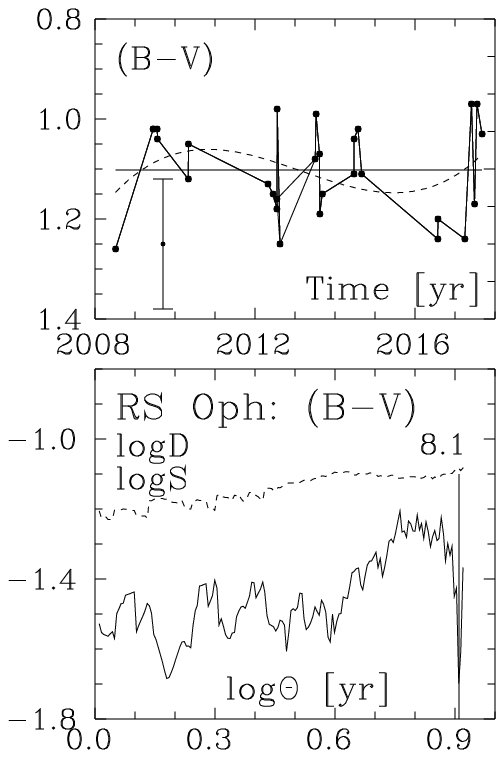, width=.3\textwidth}}
\caption[] {Application of our method on the variability of RS Oph on the scale of 10 years.
The top panels show the average magnitudes and color indexes from 29 series in 2008–2017. 
The solid horizontal lines show the common average values. The dashed curves show the optimal cubic polynomials. 
The bottom panels show the $F_D$ and $F_S$ (Eq. 4), build on the original magnitude and color data. } 
\label{fig.7}   \end{center}  \end{figure} 

\section*{4. Results}

\section*{4.1. Distributions of the time sizes $T_S$}

Using the structure functions $f_S$ (Eq.4) for 58 flickering series, 
29 in B band and 29 in V band we revealed 80 structure sizes, 
42 in B band and 38 in V band. 
We found one or two mutually independent structure sizes in 57 \% or 41 \% of the B band  and in 66 \% or 34 \% in the V band observations, 
respectively. Only once we found 3 apparently independent structures (Fig. 4).

Figure 8 juxtaposes the sizes of the detected structures $T_S$ with 
the duration of the run $T_M$ (\textit{a})and the time resolutions 
(the mean interval between two data points) $\delta T$ (\textit{b}). 

The quantity of the large structure sizes ($T_S$ > 60 min) is small  because they are detectable only in the long series. 
The shorter structure sizes ($T_S$ = 20-40 min) are numerous because they are detectable in almost all series (Fig.8(\textit{a})). 
Doubtless, numerous structures with short sizes, ($T_S \approx$ 10 min), are smeared and undetectable 
by $f_S$ (Fig.3-6). 

Nevertheless, the detectability of structures with sizes 10 – 120 min does not depend on the time interval between data points, which is 
$\delta T$ = 0.5 – 3 min (Fig. 8\textit{b}).

Figure 9 (\textit{a}) shows the distribution of the detected  structure sizes $T_S$ over linear scale. The majority of the time sizes (about 3/4) 
belong to the interval 10-40 min where the most widespread structures (about 1/5 of all) have time sizes about 20 min. Small part of all structure sizes (about 1/4) belong to the interval 60-100 min. If these structure sizes are regarded as quasi-periods, then the respective cycles per day are 144 - 36 c/d, 72 c/d and 24 – 15 c/d, respectively. By the inconsistency of long quasi-periods this distribution has approximately log-normal shape, but in principle  it might be regarded also over logarithmic scale.    

Figure 9 (\textit{b}) shows the distribution of $T_S$ in logarithmic scale. 
Now the histogram poses pronounced local maximums at about 10, 21, 36 and 74 min. 
They correspond to about 144, 69, 40 and 19 c/d, respectively. 
It seems the flickering with time size about 10 min is poorly detectable because it is not unique, 
it contains flickering with time sizes 8 - 12 min. Therefore we may  consider that the quasi-period with size 
about 21 min, corresponding to about 69 c/d, seems to ne fundamental time size in the flickering variability of RS Oph. 

\begin{figure}[!htb]  \begin{center}
\centering{\epsfig{file=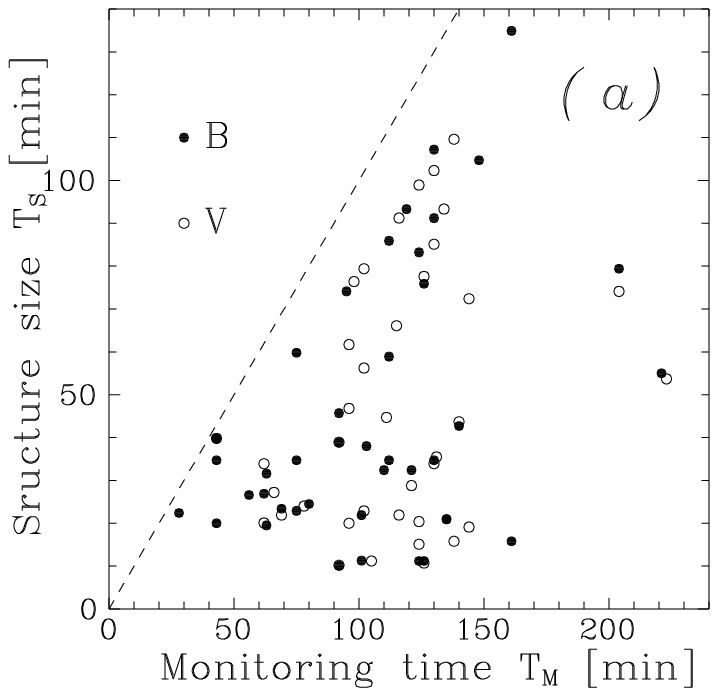, width=.4\textwidth}}
\centering{\epsfig{file=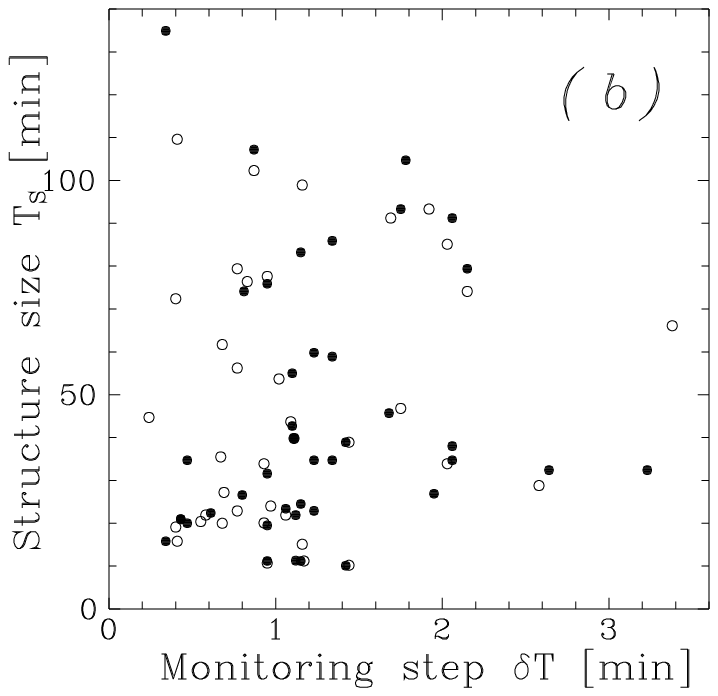, width=.4\textwidth}}
\caption[] {(\textit{a}) Juxtaposition of the monitoring times $T_M$ of the series with the sizes of the detected structures $T_S$ in B band (dotes) and in V band (circles). The dashed line with slope of 45° bounds the right-bottom zone of the possible detection. (\textit{b}) Juxtaposition of the mean time resolutions of the series $\delta T$ with the detected structure sizes $T_S$.  } 
\label{fig.8}   \end{center}  \end{figure} 

\begin{figure}[!htb]  \begin{center}
\centering{\epsfig{file=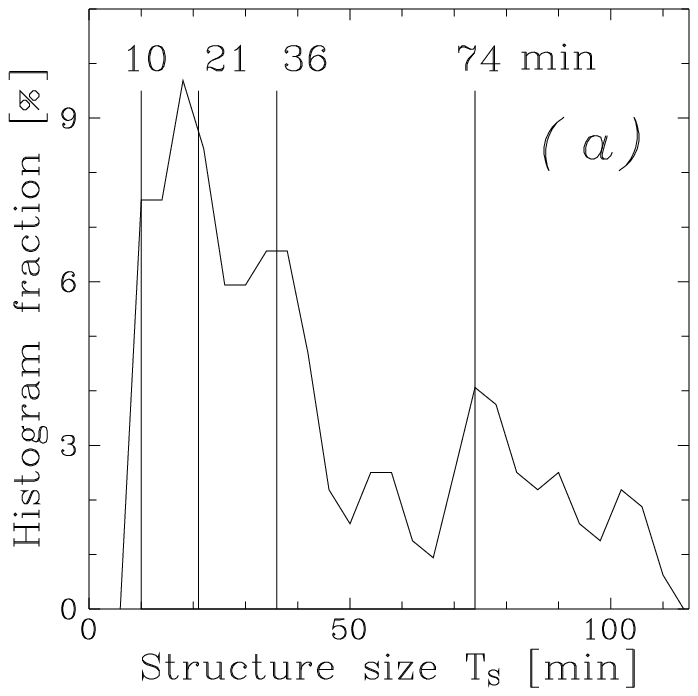, width=.4\textwidth}}
\centering{\epsfig{file=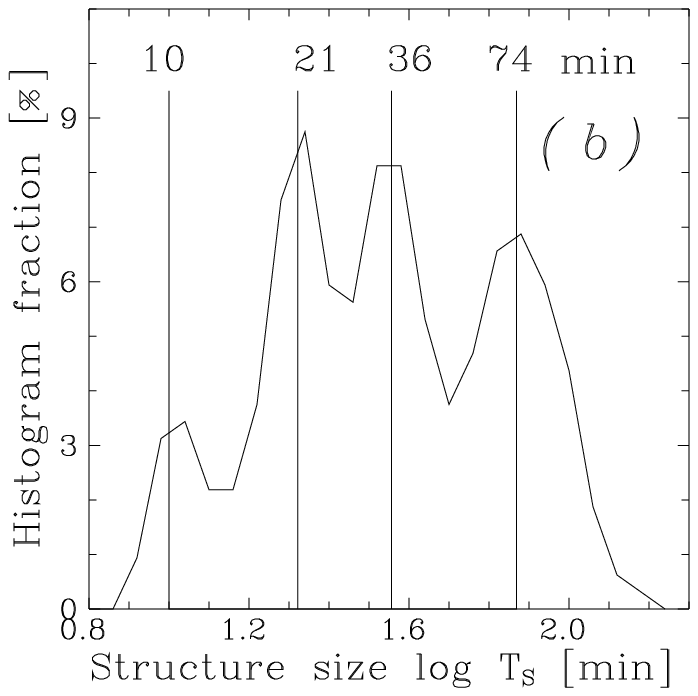, width=.4\textwidth}}
\caption[] { \textit{(a)} Distributions of the detected structure sizes $T_S$ in linear scale  with binning size 4 min  
and  (\textit{b})  in logarithmic scale with binning size 0.6 in B and V bands together.  } 
\label{fig.9}   \end{center}  \end{figure} 

\subsection*{4.2. Correlation between the time size $T_S$ and the flickering energy $E_S$}
 
Having 80 time sizes (Section 4.1) we used the deviation functions $f_D$ (Eq.4) for estimations of the magnitude standard deviations which are  placed just above the $T_S$ (see Fig.3). We interpret these standard deviations as energetic characteristics of the time structures and denote them as  $E_S$.

The "energies" $E_S$ belong to the interval 0.02 – 0.11 mag (Table 3). Because of the definition of the stellar magnitude, 
the magnitude differences $\Delta m$ in this digital range, in magnitudes,  correspond well to the relative fluxes $\Delta F/F$ 
in the same digital range, but in percentages (2 \% – 11 \%).  More accurately, the respective relative fluxes are 1.84 \% - 10.65 \% (see Georgiev, 2018).  

Figure 10 (\textit{a}) shows that the energies $E_S$ correlate moderately with 
its structure sizes log $T_S$. The coefficients of correlation are 0.68 and 0.57 in B and V bands. 
The regression slope coefficients are 0.04 and 0.03 in B and V bands, respectively.

\begin{figure}[!htb]  \begin{center}
\centering{\epsfig{file=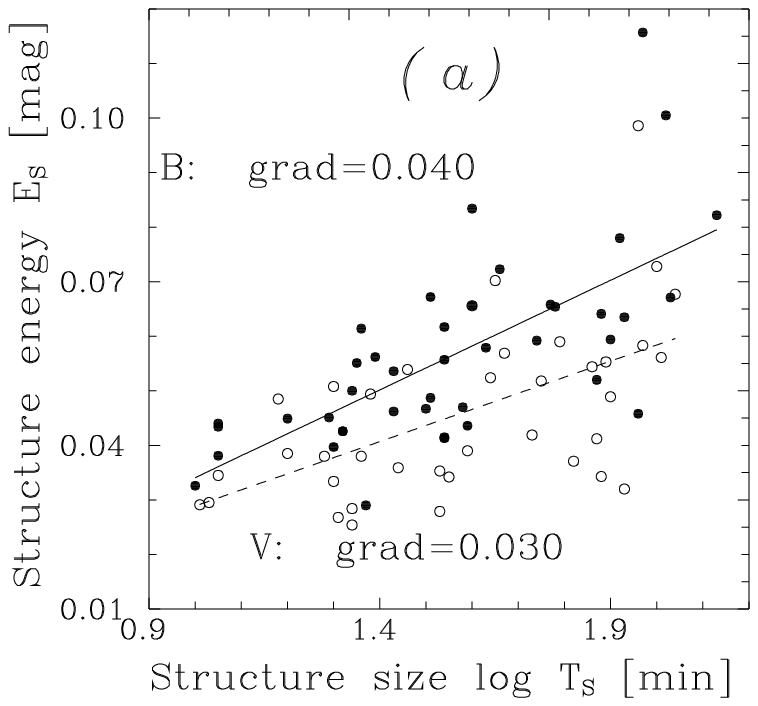, width=.4\textwidth}}
\centering{\epsfig{file=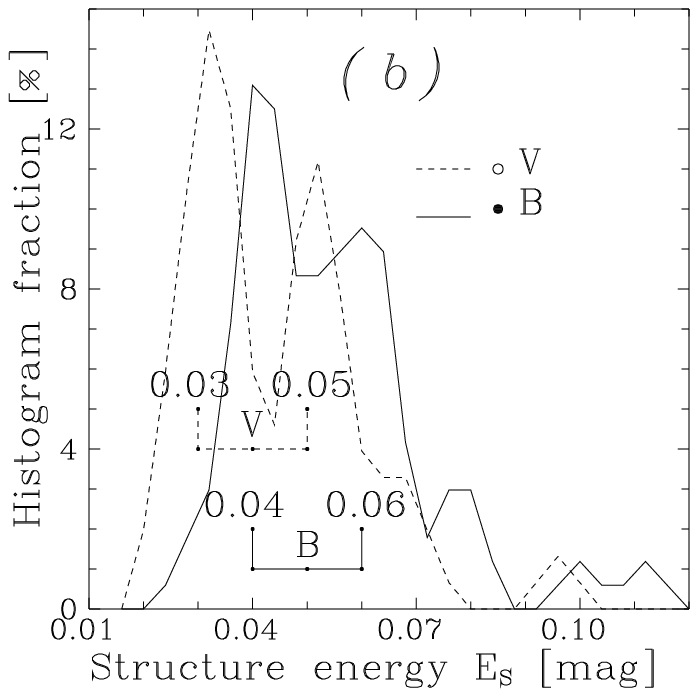, width=.37\textwidth}}
\caption[] {(\textit{a}). Correlations between the structure size log $T_S$ and its cumulative flickering energy $E_S$; (\textit{b}) Distributions of the energies $E_S$ in B and V band. } 
\label{fig.7}   \end{center}  \end{figure} 

Figure 10 (\textit{b}) shows that the energies $E_S$ pose bimodal distributions in B and V bands. 
The local maximums of the distributions are placed about 0.04 and 0.06 (4 \% and 6 \%) in B band, 
as well as about 0.03 and 0.05 (3 \% and 5 \%) in V band.  The left and right maximums may be associated 
with the structure sizes about 10 – 21 min and 37 – 74 min, respectively. 

\section*{5. Conclusions}

We applied the fractal approach for  revealing and energetic characterization of time 
structures in the apparent chaos of the flickering (Figs.3-7) of RS Oph for 58 flickering series. We found that: 

(1) in 12 \% of our flickering series, in cases of relatively low states of the system, increasing of the common brightness with mean gradient $G$ < 0.15 mag/h is registered. Opposite cases are not observed (Fig.1);  

(2) quasi-periodic structures are found in all flickering time series. The quasi-period of about 21 min (69 c/d) is the most widespread time structure size; 

(3) a system of local maximums in the distribution of the quasi-periods at about 10, 21, 36 and 74 min (Fig. 9(\textit{b})) is revealed;

(4) correlations between the structure time size, log$T_S$, and  its cumulative flickering energy, 
$E_S$, with regression slope coefficients 0.04 and 0.03 in B and V bands, respectively, are found (Fig. 10). 

In paper II, we will transform the stellar magnitudes 
into fluxes in order to apply other tools of the fractal analysis. 

\vskip 0.5cm 

{\bf Acknowledgments: }
This work was partly supported by  grants  DN 08/\\1~13.12.2016, DN 18/13~12.12.2017 (Bulgarian National Science Fund), 
and  AYA2016-76012-C3-3-P from the Spanish Ministerio de Econom\'ia y Competitividad (MINECO).

The authors thank to the referee Dr. Andreja Gomboc for the attention and recommendations to this paper. 
{}     

\newpage
\begin{table}   \centering
\caption{B band data. The columns are as follow: 
1 - designation of the flickering series,  2 - date of observation - YYYYMMDD and 200x.xxx, 3 - total monitoring time (min), 
4 - number of the monitoring points, 5 - mean time resolution of the monitoring  (min), 
6 - slope coefficient of the input flickering bias (min/h), 7 - decreasing factor of the SD after bias removing (Eq. 1), 
8 - average magnitude, 9 - magnitude SD (after unbiasing), 10 - full amplitude of the flickering (after unbiasing), 
11 - telescope.           }       
\begin{tabular}{ccr rcccccccc}    \hline    \\    
\#  &    Y  &  $T_M$  & $N_P$ &  $\delta T$ &   $G$   & $C_{SD}$  & $m_0$  &  $m_{SD}$ &  $m_A$ & Tel \\    \hline   \\
 1  &     2  &   3   &  4  &   5   &   6    &   7  &   8   &   9  &  10 & 11\\  \hline  \\
 01B & 20080706 2008.516 & 102.8 &  50 &  2.06 &  0.014 & 1.01 & 12.47 & 0.051 & 0.220 & 70cm~Sch  \\
 02B & 20090614 2008.455 &  55.9 &  70 &  0.80 & -0.012 & 1.00 & 11.73 & 0.060 & 0.253 & 60cm~Roz  \\
 03B & 20090615 2009.458 & 112.4 &  84 &  1.34 & -0.007 & 1.00 & 11.76 & 0.066 & 0.360 & 60cm~Roz  \\
 04B & 20090721 2009.556 &  80.5 &  70 &  1.15 &  0.201 & 1.60 & 12.02 & 0.061 & 0.275 & 60cm~Roz  \\
 05B & 20090723 2009.561 &  28.2 &  46 &  0.61 & -0.207 & 1.13 & 11.97 & 0.055 & 0.202 & 70cm~Sch  \\
 06B & 20100430 2010.333 &  92.5 &  65 &  1.42 & -0.052 & 1.08 & 11.74 & 0.054 & 0.238 & 60cm~Roz  \\
 07B & 20100501 2010.336 & 119.0 &  68 &  1.75 &  0.056 & 1.03 & 11.43 & 0.124 & 0.445 & 60cm~Roz  \\
 08B & 20120427 2012.325 &  74.9 &  61 &  1.23 &  0.075 & 1.09 & 12.11 & 0.064 & 0.318 & 60cm~Roz  \\
 09B & 20120613 2012.453 & 109.7 &  34 &  3.23 &  0.120 & 1.49 & 12.52 & 0.058 & 0.318 & 60cm~Roz  \\
 10B & 20120718 2012.548 & 134.9 & 317 &  0.43 & -0.018 & 1.03 & 12.72 & 0.045 & 0.239 & 70cm~Sch  \\
 11B & 20120721 2012.556 & 220.7 & 200 &  1.10 &  0.016 & 1.03 & 12.41 & 0.065 & 0.231 & 60cm~Bel  \\
 12B & 20120815 2012.624 & 148.0 &  83 &  1.78 & -0.262 & 2.14 & 12.90 & 0.101 & 0.359 & 60cm~Roz  \\
 13B & 20120816 2012.548 & 134.9 & 317 &  0.43 & -0.018 & 1.03 & 12.72 & 0.045 & 0.395 & 2.0m~Roz  \\
 14B & 20130702 2013.505 & 121.4 &  46 &  2.64 &  0.016 & 1.01 & 12.19 & 0.073 & 0.396 & 60cm~Roz  \\
 15B & 20130710 2013.527 & 139.7 & 127 &  1.10 &  0.061 & 1.27 & 12.03 & 0.054 & 0.289 & 70cm~Sch  \\
 16B & 20130812 2013.615 & 160.6 & 470 &  0.34 & -0.006 & 1.00 & 12.13 & 0.083 & 0.360 & 70cm~Sch \\
 17B & 20130813 2013.619 &  42.8 &  92 &  0.47 &  0.072 & 1.06 & 12.58 & 0.040 & 0.175 & 70cm~Sch+60cm~Roz\\
 18B & 20130906 2013.683 &  43.1 &  39 &  1.11 & -0.029 & 1.00 & 12.02 & 0.090 & 0.315 & 60cm~Bel  \\
 19B & 20140621 2014.475 &  68.7 &  65 &  1.06 & -0.019 & 1.02 & 12.54 & 0.032 & 0.148 & 60cm~Bel  \\
 20B & 20140622 2014.478 & 124.4 & 108 &  1.15 & -0.241 & 2.17 & 12.62 & 0.075 & 0.315 & 60cm~Bel  \\
 21B & 20140729 2014.578 &  62.4 &  32 &  1.95 & -0.043 & 1.04 & 12.30 & 0.047 & 0.185 & 70cm~Sch  \\
 22B & 20140831 2014.667 &  92.3 &  55 &  1.68 & -0.175 & 1.53 & 12.09 & 0.065 & 0.266 & 70cm~Sch  \\
 23B & 20160726 2016.570 & 129.7 &  63 &  2.06 & -0.130 & 2.05 & 12.30 & 0.046 & 0.222 & 70cm~Sch  \\
 24B & 20160728 2016.575 & 204.2 &  95 &  2.15 &  0.058 & 1.33 & 12.28 & 0.066 & 0.314 & 70cm~Sch  \\
 25B & 20170329 2017.245 &  95.0 & 118 &  0.81 & -0.154 & 1.71 & 12.04 & 0.051 & 0.243 & 70cm~Sch  \\
 26B & 20170528 2017.411 & 101.0 &  90 &  1.12 & -0.004 & 1.00 & 12.13 & 0.060 & 0.293 & 60cm~Bel  \\
 27B & 20170626 2017.489 & 130.0 & 150 &  0.87 &  0.005 & 1.00 & 11.68 & 0.067 & 0.258 & 60cm~Bel  \\
 28B & 20170719 2017.553 & 126.3 & 133 &  0.95 & -0.069 & 1.19 & 11.71 & 0.062 & 0.300 & 41cm~Ja\'{e}n  \\
 29B & 20170904 2017.678 &  63.0 &  66 &  0.95 &  0.107 & 1.25 & 11.68 & 0.044 & 0.189 & 41cm~Ja\'{e}n  \\
         \\      \hline\\   \end{tabular}     \end{table}
         
\newpage
\begin{table}   \centering
\caption{V band  data. 
1 - designation of the flickering series. 2 - date of observations - YYYYMMDD and 200x.xxx, 3 - total monitoring time (min), 
4 - number of the monitoring points, 5 - mean time resolution of the monitoring (min), 6 - slope coefficient of the input flickering (min/h),
 7 - decreasing factor of the SD after bias removing (Eq. 1), 8 - average magnitude, 
 9 - magnitude SD (after unbiasing), 10 - full amplitude of the flickering (after unbiasing), 
 11 - telescope.  }
\begin{tabular}{ccrrcccccccccc}    \hline    \\      
\#  &    Y  &  $T_M$  & $N_P$ &  $\delta T$ &   $G$   & $C_{SD}$  & $m_0$  &  $m_{SD}$ &  $m_A$ & Tel \\    \hline   \\       
 1  &    2   &   3   &  4  &   5   &   6    &  7   &   8   &   9   &  10 & 11 \\  \hline  \\
 01V & 20080706 2008.516 & 131.0 & 196 &  0.67 &  0.052 & 1.25 & 11.21 & 0.042 & 0.180 & 2.0m~Roz \\
 02V & 20090614 2009.455 &  65.7 &  95 &  0.69 &  0.068 & 1.12 & 10.71 & 0.043 & 0.205 & 60cm~Bel \\
 03V & 20090615 2009.458 & 102.3 & 133 &  0.77 &  0.004 & 1.00 & 10.74 & 0.053 & 0.332 & 60cm~Bel \\
 04V & 20090721 2009.556 &  78.5 &  81 &  0.97 &  0.130 & 1.22 & 11.00 & 0.076 & 0.337 & 60cm~Bel \\
 05V & 20090723 2009.561 & 111.4 & 467 &  0.24 &  0.048 & 1.07 & 10.93 & 0.065 & 0.348 & 2.0m~Roz \\
 06V & 20100430 2010.333 &  92.2 &  64 &  1.44 & -0.036 & 1.06 & 10.62 & 0.046 & 0.199 & 70cm~Sch \\
 07V & 20100501 2010.336 & 116.4 &  69 &  1.69 &  0.061 & 1.05 & 10.38 & 0.108 & 0.405 & 70cm~Sch \\
 08V & 20120427 2012.325 &  96.3 & 141 &  0.68 &  0.038 & 1.05 & 10.98 & 0.055 & 0.269 & 60cm~Bel \\
 09V & 20120613 2012.453 & 114.8 &  34 &  3.38 &  0.086 & 1.46 & 11.37 & 0.043 & 0.269 & 70cm~Sch \\
 10V & 20120718 2012.548 & 115.9 & 200 &  0.58 & -0.006 & 1.01 & 11.54 & 0.030 & 0.200 & 2.0m~Roz \\
 11V & 20120721 2012.556 & 223.5 & 220 &  1.02 &  0.016 & 1.07 & 11.43 & 0.044 & 0.151 & 60cm~Bel \\
 12V & 20120815 2012.624 & 134.3 &  70 &  1.92 & -0.241 & 2.81 & 11.65 & 0.061 & 0.224 & 70cm~Sch \\
 13V & 20120816 2012.548 & 124.0 & 224 &  0.55 &  0.034 & 1.20 & 11.56 & 0.031 & 0.272 & 70cm~Sch \\
 14V & 20130702 2013.505 & 121.4 &  47 &  2.58 &  0.006 & 1.00 & 11.11 & 0.060 & 0.253 & 60cm~Roz \\
 15V & 20130710 2013.527 & 139.7 & 128 &  1.09 &  0.062 & 1.33 & 11.04 & 0.049 & 0.219 & 70cm~Sch \\
 16V & 20130812 2013.615 & 137.6 & 335 &  0.41 &  0.021 & 1.02 & 11.06 & 0.067 & 0.289 & 70cm~Sch \\
 17V & 20130813 2013.619 & 144.1 & 362 &  0.40 &  0.049 & 1.20 & 11.39 & 0.050 & 0.252 & 2.0m~Roz \\
 18V & 20130906 2013.683 &  43.1 &  39 &  1.11 & -0.022 & 1.00 & 10.87 & 0.071 & 0.252 & 60cm~Bel \\
 19V & 20140621 2014.475 &  68.7 &  65 &  1.06 & -0.005 & 1.00 & 11.43 & 0.027 & 0.263 & 60cm~Bel \\
 20V & 20140622 2014.478 & 124.4 & 107 &  1.16 & -0.202 & 1.88 & 11.58 & 0.076 & 0.121 & 60cm~Bel \\
 21V & 20140729 2014.578 &  93.2 &  43 &  2.17 & -0.090 & 1.25 & 11.28 & 0.052 & 0.373 & 70cm~Sch \\
 22V & 20140831 2014.667 &  96.3 &  55 &  1.75 & -0.141 & 1.56 & 10.98 & 0.051 & 0.206 & 70cm~Sch \\
 23V & 20160726 2016.570 & 129.7 &  64 &  2.03 & -0.096 & 2.16 & 11.06 & 0.032 & 0.185 & 2.0m~Roz \\
 24V & 20160728 2016.575 & 204.2 &  95 &  2.15 &  0.045 & 1.37 & 11.08 & 0.048 & 0.245 & 70cm~Sch \\
 25V & 20170329 2017.245 &  97.9 & 118 &  0.83 & -0.107 & 1.72 & 10.80 & 0.035 & 0.176 & 70cm~Sch \\
 26V & 20170528 2017.411 & 105.0 &  90 &  1.17 & -0.012 & 1.01 & 11.16 & 0.051 & 0.237 & 60cm~Bel \\
 27V & 20170626 2017.489 & 129.8 & 150 &  0.87 & -0.001 & 1.00 & 10.51 & 0.056 & 0.231 & 60cm~Bel \\
 28V & 20170719 2017.553 & 126.2 & 133 &  0.95 & -0.059 & 1.19 & 10.74 & 0.054 & 0.252 & 41cm~Ja\'{e}n \\
 29V & 20170904 2017.678 & 62.1 &  67 &  0.93 &  0.121 & 1.48 & 10.65 & 0.034 & 0.152\ & 41cm~Ja\'{e}n \\
         \\      \hline\\   \end{tabular}     \end{table}
                  
\newpage
\begin{table}   \centering
\caption{ Output results.
1 - designation of the B flickering series, 2 - date of the observations, 3 - time structure size in B band (min),
4 - logarithm of the time structure size, 5 - the respective energy, in magnitudes,      
6 - designation of the V flickering series, 7 - date of the observations, 8 - time structure size in V band (min),
9 - logarithm of the time structure size, 10 - the respective energy, in magnitudes. See also the end of Section 2.  }
\begin{tabular}{cccccc | cccccccc} \hline \\                               
\#B & Date & $T_S$(B) & log $T_S$ & $E_S$ & & & \#V  &   Date   & $T_S$(B)& log $T_S$&  $E_S$ \\    \hline   \\
  1 &  2   &  3       &  4        &  5    & & &  6  &     7    &  8   &  9  &  10 \\      \hline   \\
 01B & 20080706 & 38.0 & 1.58 & 0.0470 &  &  &  01V & 20080706 & 35.5 & 1.55 & 0.0342\\
 02B & 20090614 & 26.6 & 1.43 & 0.0536 &  &  &  02V & 20090614 & 27.2 & 1.44 & 0.0359\\
 03B & 20090615 & 34.7 & 1.54 & 0.0557 &  &  &  03V & 20090615 & 22.9 &1.36 & 0.0380\\
 03B & 20090615 & 58.9 & 1.77 & 0.0658 &  &  &  03V & 20090615 & 56.2 &1.75 & 0.0518\\
 03B & 20090615 & 85.9 & 1.93 & 0.0635 &  &  &  03V & 20090615 & 79.4 &1.90 & 0.0489\\
 04B & 20090721 & 24.5 & 1.39 & 0.0562 &  &  &  04V & 20090721 & 24.0 & 1.38 & 0.0494\\
 05B & 20090723 & 22.4 & 1.35 & 0.0551 &  &  &   -  &	  -    & -    & -    & - \\
 -   &    -     &   -  &      &  -     &  &  &  05V & 20090723 & 44.7 & 1.65 & 0.0702\\
 06B & 20100430 & 10.1 & 1.00 & 0.0326 &  &  &  06V & 20100430 & 10.2 & 1.01 & 0.0291\\
 06B & 20100430 & 38.9 & 1.59 & 0.0436 &  &  &   -  &	-     & -     & -   &  - \\
 07B & 20100501 & 93.3 & 1.97 & 0.1157 &  &  &  07V & 20100501 & 91.2 & 1.96 & 0.0986\\
 08B & 20120427 & 22.9 & 1.36 & 0.0614 &  &  &  08V & 20120427 & 20.0 & 1.30 & 0.0508\\
 08B & 20120427 & 34.7 & 1.54 & 0.0617 &  &  &   -  &	 -     &   -  &     &  - \\
 08B & 20120427 & 59.8 & 1.78 & 0.0654 &  &  &  08V & 20120427 & 61.7 & 1.79 & 0.0590\\
 09B & 20120613 & 32.4 & 1.51 & 0.0487 &  &  &  -   &	  -   &  -   &-    &  - \\
     &     -    &   -  &   -  &    -   &  &  &  09V & 20120613 & 66.1 &1.82 & 0.0371\\
 10B & 20120718 & 21.0 & 1.32 & 0.0426 &  &  &  10V & 20120718 & 21.9 &1.34 & 0.0284\\
 11B & 20120721 & 55.0 & 1.74 & 0.0592 &  &  &  11V & 20120721 & 53.7 &1.73 & 0.0419\\
 12B & 20120815 &104.7 & 2.02 & 0.1005 &  &  &  12V & 20120815 & 93.3 &1.97 & 0.0583\\
 13B & 20120816 & 20.9 & 1.32 & 0.0426 &  &  &  13V & 20120816 & 20.4 &1.31 & 0.0268\\
 14B & 20130702 & 32.4 & 1.51 & 0.0672 &  &  &  14V & 20130702 & 28.8 &1.46 & 0.0539\\
 15B & 20130710 & 42.7 & 1.63 & 0.0579 &  &  &  15V & 20130710 & 43.7 &1.64 & 0.0524\\
 16B & 20130812 & 15.8 & 1.20 & 0.0449 &  &  &  16V & 20130812 & 15.8 &1.20 & 0.0385\\
 16B & 20130812 &134.9 & 2.13 & 0.0822 &  &  &  16V & 20130812 &109.6 &2.04 & 0.0677\\
 17B & 20130813 & 20.0 & 1.30 & 0.0397 &  &  &  17V & 20130813 & 19.1 &1.28 & 0.0380\\
 17B & 20130813 & 34.7 & 1.54 & 0.0415 &  &  &   -  &	 -    & -    & -    & - \\
  -  &     -    &   -  &   -  &        &  &  &  17V & 20130813 & 72.4 &1.86 & 0.0544\\
 18B & 20130906 & 39.8 & 1.60 & 0.0834 &  &  &  18V & 20130906 & 39.8 &1.60 & 0.0656\\
 19B & 20140621 & 23.4 & 1.37 & 0.0290 &  &  &  19V & 20140621 & 21.9 &1.34 & 0.0254\\
 20B & 20140622 & 11.2 & 1.05 & 0.0440 &  &  &  20V & 20140622 & 15.1 &1.18 & 0.0485\\
 20B & 20140622 & 83.2 & 1.92 & 0.0780 &  &  &  20V & 20140622 & 98.9 &2.00 & 0.0728\\
 21B & 20140729 & 26.9 & 1.43 & 0.0462 &  &  &  21V & 20130729 & 27.5 &1.44 & 0.0454\\
 22B & 20140831 & 45.7 & 1.66 & 0.0723 &  &  &  22V & 20140831 & 46.8 &1.67 & 0.0569\\
 23B & 20160726 & 34.7 & 1.54 & 0.0413 &  &  &  23V & 20160726 & 33.9 &1.53 & 0.0279\\
 23B & 20160726 & 91.2 & 1.96 & 0.0458 &  &  &  23V & 20160726 & 85.1 &1.93 & 0.0320\\
 24B & 20160728 & 79.4 & 1.90 & 0.0594 &  &  &  24V & 20160728 & 74.1 &1.87 & 0.0412\\
 25B & 20170329 & 74.1 & 1.87 & 0.0520 &  &  &  25V & 20170329 & 76.4 &1.88 & 0.0343\\
 26B & 20170528 & 11.3 & 1.05 & 0.0434 &  &  &  26V & 20170528 & 11.2 &1.05 & 0.0345\\
 26B & 20170528 & 21.9 & 1.34 & 0.0500 &  &  &   -  &	  -   &  -   & -    &  - \\
   - &    -     &  -   &  -   &   -    &  &  &  28V & 20170719 & 10.7 &1.03 & 0.0295\\
 27B & 20170626 &107.2 & 2.03 & 0.0671 &  &  &  27V & 20170626 &102.3 &2.01 & 0.0561\\
 28B & 20170719 & 11.2 & 1.05 & 0.0381 &  &  &   -  &	 -   &  -   &  -  &    - \\
 28B & 20170719 & 75.9 & 1.88 & 0.0641 &  &  &  28V & 20170719 & 77.6 &1.89 & 0.0553\\
 29B & 20170904 & 19.5 & 1.29 & 0.0451 &  &  &  29V & 20170904 & 20.1 &1.30 & 0.0334\\
 29B & 20170904 & 31.6 & 1.50 & 0.0467 &  &  &  29V & 20170904 & 33.9 &1.53 & 0.0353\\
    \\      \hline\\   \end{tabular}    \end{table}              
                  
\end{document}